
\input phyzzx
\input epsf
\def\CR{C_{2R}}
\def\tn{\tilde n}
\def\gam{\gamma}
\def\ws{world-sheet}
\def\diff{diffeomorphism}
\def\diffs{diffeomorphisms}
\def\cF{{\cal F}}
\def\Imt{{\rm Im}\tau}
\def\ts{target space}
\def\om{\omega}
\def\mod{{\rm \ mod}}
\def\ram{ramification}
\def\NG{Nambu-Goto}
\def\figcap#1#2{{\singlespace\tenpoint\noindent
{\bf Figure #1}\ #2}
\vskip.2in}
\def\tablecap#1#2{{\singlespace\tenpoint\noindent
{\bf Table #1}\ #2}
\vskip.2in}
\Pubnum={UVA-HET-92-10\cr
hepth@xxx/9301003}
\date={December 1992}
\pubtype={}
\titlepage
\title{Summing Over Inequivalent Maps in the String Theory Interpretation
\break
of Two Dimensional QCD}
\bigskip
\author {
Joseph~A.~Minahan\footnote\star
{minahan@gomez.phys.virginia.edu}}
\address{Department of Physics,  Jesse Beams Laboratory,\break
University of Virginia, Charlottesville, VA 22901 USA}
\bigskip
\abstract{
Following some recent work by Gross, we consider the partition
function for QCD on a two dimensional torus and study its stringiness.  We
present strong evidence that the free energy corresponds to a sum over
branched surfaces with small handles mapped into the target space.
The sum is modded out by all diffeomorphisms on the world-sheet.
This leaves a sum over disconnected classes of maps.
We prove that the free energy gives a consistent result for all smooth
maps of the torus into the torus which cover the target space $p$ times,
where $p$ is prime, and conjecture that this is true for all coverings.
Each class can also contain integrations over the positions of
branch points and small handles which act as ``moduli'' on the surface.
We show that the free energy is consistent for any number of handles
and that the first few leading terms are consistent with contributions
from maps with branch points.
}
\endpage

\def\NP{{\it Nucl. Phys.\ }}
\def\PL{{\it Phys. Lett.\ }}
\def\PRD{{\it Phys. Rev. D\ }}

\def\PRL{{\it Phys. Rev. Lett.\ }}
\def\CMP{{\it Comm. Math. Phys.\ }}

\def\MPL{{\it Mod. Phys. Lett. A\ }}

\def\ZETF{{\it Zh. Eksp. Teor. Fiz.}}

\REF\Poly{A.~Polyakov, \PL {\bf 103B} (1981) 211.}
\REF\DZ{S.~Deser and B.~Zumino, \PL {\bf65B} (1976) 369.}
\REF\BDH{L.~Brink, P.~Di~Vecchia and P.~Howe, \PL {\bf 65B} (1976) 471.}
\REF\Friedan{D.~Friedan, in {\it Recent Advances in Field Theory and
Statistical Mechanics}, eds. J.~Zuber and R. Stora.  Proc. of 1982 Les Houches
Summer School, p. 839.}
\REF\Alvarez{O.~Alvarez, \NP {\bf B216} (1983) 125.}
\REF\Durh{B.~Durhuus, P.~Oleson and J.~Petersen, \NP {\bf B198} (1982) 157.}
\REF\Fuji{K.~Fujikawa, \NP {\bf B226} (1983) 437.}
\REF\Gross{D.~Gross, LBL and Princeton preprints LBL 33233, PUPT 1356; LBL
33232
PUPT 1355, 1992.}
\REF\tHooft{G.~'t~Hooft, \NP {\bf B75} (1974) 461.}
\REF\CCG{C.~Callan, N.~Coote and D.~Gross, \PRD {\bf13} (1976) 1649.}
\REF\BBHP{W.~Bardeen, I.~Bars, A.~Hanson and R.~Peccei, \PRD {\bf13} (1976)
2364.}
\REF\BarsI{I.~Bars, \PRL {\bf36} (1976) 1521; \NP {\bf B111} (1976) 413.}
\REF\Migdal{A.~Migdal, \ZETF {\bf 69} (1975) 810.}
\REF\Rus{B.~Rusakov, \MPL {\bf5} (1990) 693.}
\REF\GSW{M.~Green, J.~Scharz and E.~Witten, {\it Superstring Theory}, Cambridge
University Press, 1987.}
\REF\Polch{J.~Polchinski, \CMP {\bf 104} (1986) 37.}
\REF\FK{H.~Farkas and I.~Kra, {\it Riemann Surfaces}, Springer-Verlag, 1980.}
\REF\GrTay{D.~Gross and W.~Taylor, {\it to appear.}}

\chapter{Introduction}

String theory has progressed tremendously since its beginning in the Dual
Model Days of the 1960's.  However, the direction of this progression
took an unexpected turn about twenty years ago.  Instead of string theory
being used as a tool to describe strong interactions, it has since
served as a means to unify all forces in Nature.

Since the turn away from strong interactions, there have been many
technical achievements in string theory.
Among the most notable achievements are those of Polyakov [\Poly] and others
[\BDH-\Fuji],
who realized that string amplitudes are given by integrations over all
geometries of two dimensional punctured surfaces.
In other words, string theory is basically two dimensional quantum gravity
coupled to matter fields.
The path integral is given by the sum over all metrics modded out by
diffeomorphisms.  If the strings are critical, then the path integral
can be modded out by conformal transformations as well.  This then leaves
an integral over the moduli of the surface, a finite dimensional space.

But for many reasons, this theory is not QCD.  However, QCD still looks very
stringy, at least in the confining phase.  Hence a natural question to ask is
to
what extent do the ideas of Polyakov string theory apply to the strong
interactions.

Gross has recently proposed a nice way to start probing this question [\Gross].
His idea is to study matter-free QCD in two dimensions and to determine
the stringiness of this particular theory.
The great advantage of two dimensions is that the theory is solvable.
Hence one can analyze the solutions and decide if they look stringy or not.
If this is a string theory then one should be able to interpret the QCD
free energy as a sum over maps of two dimensional surfaces into a two
dimensional \ts.
Of course two dimensional QCD is almost a trivial theory and it is not quite
clear if everything one learns from it can be applied to the four
dimensional case.  But there might be some general principles that can
be extracted from the two dimensional case that are applicable in four
dimensions.  In particular, this might lead to a consistent formulation
of the measure for the string path integral.

This approach differs from earlier attempts to interpret two dimensional
QCD as a string theory [\tHooft-\BarsI],
in that it leaves out the quark fields.  The
resulting theory is completely trivial if the euclidean
space is flat and noncompact.  However, if the space is compact and
topologically nontrivial, then the partition function will have an
interesting structure.  It is partition functions of this type
that Gross proposes to explore.

In section 2 we review Gross' work on two dimensional QCD as a string theory.
In section 3 we carry this work out further.  We argue that for a toroidal
target space, the QCD solutions describe the equivalence classes under
\diffs\ of smooth maps into the \ts.  However, unlike Polyakov string theory,
there is no integration over \ws\ metrics.  We also present evidence that the
QCD solutions allow for branched surfaces by showing that this is consistent
with lower order terms in the perturbative expansion.  We also argue that the
solutions imply the existence of pinched handles and tubes on the surfaces.
In section 4 we close with a few remarks.

\chapter{Gross' Picture of 2d QCD}

The particular model that Gross has in mind is a lattice formulation of
QCD with a heat kernal action.  This model was recently solved by Migdal
and Rusakov [\Migdal,\Rus] and its partition function for $SU(N)$ is given by
$$Z=\sum_{\rm reps}(d_r)^{2-2G}\exp(-Ag^2C_{2R}/N),\eqn\partfun$$
where the sum is over all represenations of $SU(N)$, $A$ is the area of
the surface,
$G$ is the genus of the surface, $g/\sqrt{N}$ is the QCD coupling,
$d_R$ is the dimension of the representation and $\CR$ is the quadratic
casimir of the representation.

The representations can be summarized by a Young Tableau.  The tableau are
described by $m$ rows, with $n_i$ boxes in row $i$, which satisfy $n_i\ge n_j$
if $i<j$.  The quadratic casimir for a particular representation is given by
$$\CR={N\over2}(n+{\tn\over N}-{n^2\over N^2}),\eqn\quadcas$$
where
$$n=\sum_{i=1}^mn_i,\qquad\qquad\tn=\sum_{i=1}^mn_i(n_i-2i+1).\eqn\ntndef$$
The partition function only depends on the quantities, $N$, $G$ and the
combination $Ag^2$.  Naturally, $1/N$ acts as the string coupling and $g^2$
as the string tension.

The important quantity is the free energy, $-\log Z$.  In Polyakov string
theory, this is given as a sum over all connected Riemann surfaces,
summing over all moduli of the surface and all matter fields that live on
the world-sheet.  Each term in the sum is weighted by $(g_s)^{2\gam-2}$,
where $g_s$ is the string coupling and $\gam$ is the genus of the world-sheet.
Thus if 2d QCD is to be a string theory, then, at least perturbatively,
we should expect the free energy to be comprised of even powers of $g_s=1/N$.
Under this interpretation, the free energy is given by maps of the world-sheet
of genus $\gam$ into the target space of genus $G$.

Gross has given a beautiful demonstration of why this picture of 2d QCD is
correct [\Gross].
Supposing that the map of the world-sheet into the target space
is continuous, that is the surface has no tears, then at the very least, the
genus of the world-sheet $\gam$, must be greater than or equal to $G$.
Now consider $d_R$, which for a representation that has $n$ boxes in the
tableau, behaves as $d_R\sim N^n$ when $N>>n$.  Hence, if $G>1$ then
the partition function is dominated by the representations with a small
number of boxes.  If we approximate $\CR$ as $Nn/2$, then the free energy
can be approximated by
$$F=-\sum_{i=1}^n c_i \left({1\over N}\right)^{2(G-1)n}\exp(-nAg^2/2),
\eqn\freeen$$
where $c_i$ are constants.  Gross has interpreted this as follows:
each term in the sum represents a map from a world-sheet of genus $\gam$
which is an $n$-fold covering of the target space.  The string action
is basically the \NG\ action, the area swept out by the world-sheet multiplied
by the string tension, which
in this case is $nAg^2$.  However, there is the caveat that the world-sheet is
not
allowed to fold back on itself, otherwise there would be terms in the sum
corresponding to world-sheets whose area is not an integer multiple of $A$.
Hence the string action should contain terms that suppress the folds.
The first term in the sum has a factor of $N^{2-2G}$, which
corresponds to a world-sheet with genus $\gam=G$.  Hence, we find that there
is no contribution to the world-sheet sum until the genus is large enough so
that there can be a smooth map into the target space.  Moreover, Gross
has pointed out that if the world-sheet covers the target space $n$ times,
then the genus of the \ws\ must satisfy
$$\gam-1\ge n(G-1).\eqn\grosscon$$
This is clearly satisfied by \freeen.

Gross has also observed that the $1/N$ corrections in $\CR$ lead to terms
in the free energy with factors of $A/N$.  He has conjectured that
such terms arise from branch points or small handles on the surface.  Surfaces
with such points will have a larger genus, hence the factors of $1/N$.
Intergrating over the positions of these points gives the factors of $A$.

The string theory is described by more than just the action. One also needs
to determine the measure.
The \NG\ action is invariant under \diffs, that is, reparameterizations of the
\ws\ coordinates, thus one should expect the string
functional to
be modded out by all \diffs.  This will greatly reduce the integration
over maps of the \ws\ into the \ts.  This string theory should not
contain integrations over a \ws\ metric either.   It does not appear in the
\NG\
action, and there is otherwise no reason to introduce it.  Hence, we
should only consider a fixed \ws\ metric which will be used to
define the functional measure.
Choosing the \ws\ metric to be the \ts\ metric leads to the area factors in
the free energy.  In the next section we will further see that this leads
to consistent results.

\chapter{Diffeomorphisms, Branches, Handles and Moduli}

In this section we concentrate on the coefficients that appear in \freeen\
and on the higher order corrections to the quadratic casimirs. We will
see that the coefficients count all maps of surfaces that are not
connected by \diffs.  We will also see that the $\tn/N$ and the $n^2/N^2$ terms
in $\CR$ can be interpreted as contributions from branched surfaces
with handles.  For what follows, we will restrict our attention to $G=1$.

If we continue to approximate $\CR$ as $Nn/2$, then the free energy density
is given by
$$\cF=-{1\over A}\log\sum_{\rm reps}\exp(-Ag^2n_R/2).\eqn\freeden$$
Gross has shown this to be equal to
$$\cF=-{1\over A}\log\eta(\exp(-Ag^2/2)),\eqn\freedeneta$$
where $\eta(q)$ is Ramanujan's partition function,
$$\eta(q)=\prod_{n=1}{1\over1-q^n}.\eqn\raman$$
Hence
$$\eqalign{\cF&={1\over A}\sum_{n=1}^\infty\log(1-q^n)\cr
&=-g^2\sum_{n=1}^\infty{S_n\over nAg^2}\exp(-Ag^2n/2),}\eqn\freepart$$
where $S_n$ is the sum over positive integers that are divisors of $n$,
$$S_n=\sum_{q|n}q.\eqn\Sneq$$
Hence, for $n=2$, $S_n=3$, since $1$ and $2$ are the divisors of $2$.

In the usual Polyakov string theory, one calculates the free energy density
by summing over all possible metrics, modded out by all \diffs.  However,
the free energy density is basically a zero point amplitude.  Since the
torus has a \diff\ that is a conformal killing vector corresponding to
constant translations on the surface, then modding out by the \diffs\
requires us to divide by the area of the surface $\Imt$ [\Polch].

Let us assume then that for the QCD case, the partition function is just
a sum over all smooth maps into the torus modded out by \diffs.  For $G=1$,
it is possible to have an $n$-fold covering with $\gam=1$,  so we will
assume that every term in \freepart\ is from the \ws\ with the topology
of a torus.  We need to decide what measure to use when summing over the
maps.  This will be especially important when we consider  handles and
branches.  But it is also important for modding out the
constant translations.  The natural choice is to use the measure that exists
on the \ts\ and pull it back to the \ws.  As in Polyakov string theory, the
scale is determined by the string tension $g^2$,
and thus defines the unit of area.
Therefore, every contribution to
the torus partition should be divided by $nAg^2$ because
of the \ws\ translational invariance.  This then accounts for the denominators
in \freepart.

It would seem that modding out all maps by \diffs\ would leave only one
map, since the \ts\ fields have two degrees of freedom, the same number
as the space of \diffs.
This is true for $n=1$, but false for the higher values.  It turns out that
for these values of $n$, there are maps that cannot be continuously changed
from one to the other, but are not discrete \diffs\ of each other either.
Let us assume that the target space
is parameterized by two
vectors $(X_1,X_2)$, which define the two independent windings on the
\ts\ surface.  Likewise, let the \ws\ be parameterized by two vectors
$(\om_1,\om_2)$, which define the nontrivial windings on its surface.
We call this the winding map.
Because of reparameterization invariance on the \ws,
the winding vectors can be redefined as
$$(\om_1+\om_2,\om_2),\qquad {\rm or}\qquad (\om_2,-\om_1).\eqn\winding$$
The mapping of the \ws\ to the \ts\ maps the winding vectors $(\om_1,\om_2)$
to the winding vectors $(X_1,X_2)$.  If the map is an $n$-fold covering of the
surface, then $(\om_1,\om_2)$ will map to multiple numbers of $X_1$, $X_2$ or
both.  For example, a double covered map might be described by $(2X_1,X_2)$.
Any map that can be manipulated to this form using the operations in
\winding\ is equivalent.  It is then easy to see that there are three
independent ways to double cover the torus, $(2X_1,X_2)$, $(X_1,2X_2)$ and
$(X_1+X_2,X_1-X_2)$.  Figure 1 shows these three independent coverings.
Checking \Sneq, we find that $S_2=3$.  This suggests
that $S_n$ counts the number of independent maps of the torus into
the torus.

{\epsfxsize=6.5in\epsfbox{qcdfig1.eps}}

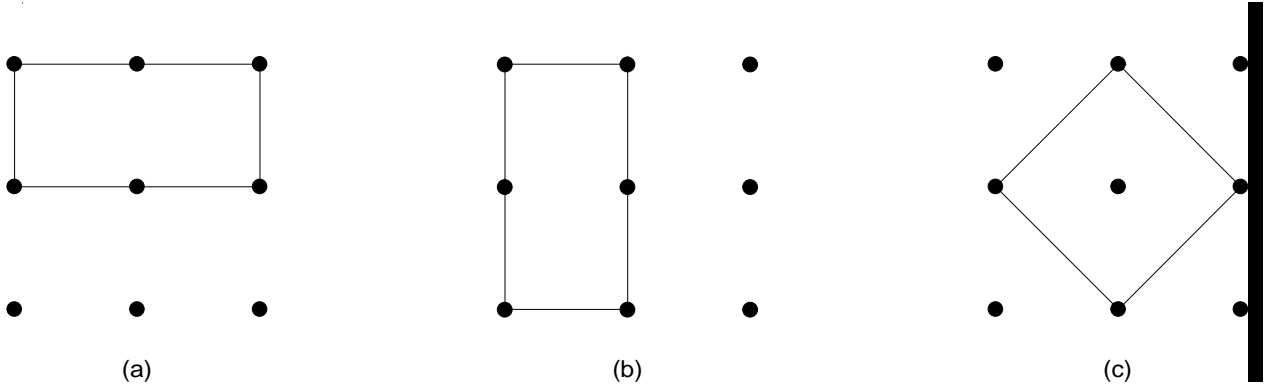
\figcap{1}{Three distinct toroidal world-sheets that
double cover the target space.  The dots represent the lattice of the toroidal
\ts.}

We now give a simple proof that $S_p$ does count the maps for $p$ prime.
Given that the area of the \ts\ is $A$, then the area of the \ws\ whose winding
map is given by $(aX_1+bX_2,cX_1+dX_2)$, is $(ad-bc)A$.
$ad-bc$ is invariant under the transformations in \winding.
Suppose that $ad-bc=p$, where $p$ is prime.  Let us further suppose
that none of the integers $a$, $b$, $c$ or $d$ are zero.  Then there are
two possible scenarios.  Either none of these integers are divisible
by $p$, or two of them are.

Let us consider the first case.
Suppose that $d>b>0$, which one can always impose using the operations in
\winding.  Since $p$ is prime, $d$ and $b$ must be relatively prime.
(That is the only common factor is 1).  Next perform
the \diff\ $(\om_1,\om_2)\to(\om_1,\om_2-\om_1)$ $n$ times, such that
$d-nb$ is as small as possible but greater than $0$.  The new winding map
is
$$(aX_1+bX_2,(c-na)X_1+(d-nb)X_2),\eqn\newmap$$
where $d-nb\equiv d'\leq b$, with an equality only if $b=1$.
Since $b$ and $d$ are relatively prime, then by construction, $b$ and $d'$
are also relatively prime.  If $d'=1$, then do the operation
$$(\om_1,\om_2)\to(\om_1-\om_2,\om_2)\eqn\diffII$$
$b$ times, leaving the winding map $(pX_1,(c-na)X_1+X_2)$.
If $d'\ne1$ then carry out the
\diff\ in \diffII\ $m$ times such that $b'=b-md'$ is as small as possible
but greater than zero.  This gives the new map
$$(a'X_1+b'X_2,c'X_1+d'X_2),\eqn\newmapII$$
where $0<d'<d$, $0<b'<b$ and $d'$ is relatively prime with $b'$.
We then repeat the process until we are left with the map
$$(pX_1,qX_1+X_2).\eqn\finalmap$$
$q$ cannot be a multiple of $p$, since this would mean that
$a$ and $c$ were multiples of $p$.
Of course, $q$ can be adjusted such that $0<q<p$, by acting
with the \diff\ $\om_2\to\om_2+\om_1$ enough times.  But clearly two maps
in the form \finalmap\ cannot be connected by a \diff\ if $q_1\ne q_2\mod p$.
Furthermore, the original map could have been transformed into the map
$(X_1+q'X_2,pX_2)$.  Hence, every map of this form is equivalent to one in
the form \finalmap.  Thus we find that there are $p-1$ distinct maps
that can be found from the first class of maps.

Turning to the second case, if $a$ and $b$ are the two integers that are
divisible by $p$, the \diff\ $\om_1\to\om_1+\om_2$ will lead to a map where
none of the integers are divisible by $p$.  A similar argument holds for $c$
and $d$ divisible by $p$.  Hence this subclass will not lead to new maps.
However, if $a$ and $c$ are divisible by $p$, then they will remain that way
under any \diff.  Furthermore, $b$ and $d$ must be relatively prime.  Hence,
using the previous argument we can find a \diff\ that sets $c$ to zero, giving
the map $(pX_1,X_2)$.  Likewise, if $b$ and $d$ are divisible by $p$,
then they stay that way under \diffs\ and the map is equivalent to
$(X_1,pX_2)$.  Therefore, combining all possible scenarios
we find a total of $p+1$ distinct maps.  Examining
\Sneq, we see that this is precisely $S_p$ if $p$ is prime.

As for non-prime $n$, we have checked the first few values of $S_n$ and have
found that they agree with the number of distinct $n$-fold maps.
Hence, we conjecture that this is true for all $n$.

Let us now examine the higher order corrections to  $\CR$ and consider their
full implications.  We can get a hint to what they might mean by realizing
that Gross' condition \grosscon\ is actually a consequence of the
Riemann-Hurwitz relation [\FK]
$$\gam-1=n(G-1)+B/2,\eqn\RiemHur$$
where $B$ is the branching number.  The branching number of a map is the
sum of the branching numbers for each point.  At each point $p$,
one can find a local coordinate $z$ which is mapped
onto another local coordinate $w$ on the other Riemann surface.
Under the map, $w$ is given by
$w=z^n$.  The branching number for this point is $b(p)=n-1$.
A point with nonzero $b(p)$ is called a ramification point and its image on
the \ts\ is called a branch point.

{}From \RiemHur\ we immediately see that the branching number is even.
Therefore,
if we only consider maps with ramification points whose branching number is
one, then the number of \ram\ points is even.  We propose that the $\tn/N$
terms in the quadratic casimirs are somehow related to these points.
At a \ram\ point of the
\ws\ the map will locally double cover the \ts.  Hence if there are
any such points at all, the entire target space must be at least double
covered.
Since the single covered target space does not have these points, we should
expect $\tn=0$ for $n=1$, which is in fact the case.

The first nonzero values for $\tn$ occur at $n=2$.
In this case, $\tn=2,-2$ for the two representations, and therefore, the
contribution to the partition function is
$$\exp(-{Ag^2\over2}(2-4/N^2))\bigl[\exp(-Ag^2/N)+\exp(Ag^2/N)\bigr].
\eqn\partexp$$
Notice that \partexp\ is an even function of $1/N$, hence the contribution to
the perturbative string expansion has only even powers of the string coupling.
In fact, the same is true for all $n\le N$, since for every Young tableau
with $n_1\le N$, there exists a transposed tableau whose value of $\tn$ has the
opposite sign.  This is illustrated in Figure 2, which shows a tableau
and its transpose.  $\tn$ is twice the sum of the numbers that appear in the
boxes.

\vskip.2in
{\epsfysize=1.8in\epsfbox{qcdfig2.eps}}

\figcap{2}{Young tableau for a representation and its
transpose.  $\tn$ is given by twice the sum of numbers in the boxes.}

Expanding the term inside the square brackets in \partexp, we find
$$2+{1\over N^2}(Ag^2)^2+{2\over N^44!}(AG^2)^4+...$$
and therefore the free energy density has the contribution
$${1\over A}\left({3\over2}+{1\over N^2}(Ag^2)^2+{2\over N^44!}(Ag^2)^4+
...\right)\exp(-Ag^2)\eqn\doublefree$$
from terms that double cover the target space.  The Riemann-Hurwitz relation
\RiemHur, suggests that the $1/N^2$ term can be
attributed to the contribution of a surface with two ramification points.  The
points connect two otherwise disconnected \ws s.  Each additional \ram\ point
leads to another factor of $1/N$.  Each
point also has a factor of $Ag^2$ associated with it because a surface with the
points at new positions
corresponds to a new \ws\ which is not connected to the
old one by a \diff. Hence, it is necessary to integrate over all positions
of the \ram\ points.  These positions are basically the
``moduli'' of the surface.  Pulling back the metric of the \ts, we find that
every \ram\ point leads to a factor of $Ag^2$, the area of the target
space multiplied by the string tension, up to symmetry factors.

The symmetry factors are as follows.  There is a factor of $1/2$ for
the two world-sheets and a factor of $1/n!$ for a surface with
$n$ \ram\ points, since they are indistinguishable.  There is also
a factor of $4$ which comes from the cuts that connect the
branch points on the \ts.
A cut joining the two points could wind either way around
the cycles of the torus.  One cut cannot be deformed into the other, hence
we find a factor of $2$ for each cycle, or a factor of $4$ altogether.
If there are more than two branch points, there is only an overall factor
of $4$, and not $4$ for each pair of points, because all of these possible
cuts can be continuously deformed into one of four types.  Putting
these factors together, the total symmetry factor for $n$ points is
$2/n!$, which agrees with \doublefree.

If the \ws\ covers the \ts\ three or more times then the counting becomes
quite complicated.  This is because many of the possible \ws s will be
equivalent to each other but determining the equivalence
is rather arduous.  We have managed to work out the factors for
a triple covering of the \ts\ with two \ram\ points and have found
agreement with the result from the free energy.  In this case, the
\ram\ points connect a surface that covers the \ts\  once with
a surface that covers the \ts\ twice.  There are three representations
that have three boxes in the tableau, and their respective values of $\tn$
are $6,0,-6$.  Plugging these values into the free energy and expanding
in $1/N$, we find that the contribution to the free energy density for a
surface
that triple covers the \ts\ and with two \ram\ points is
$${1\over A}{1\over N^2}8(Ag^2)^2\exp(-3Ag^2/2).\eqn\threedoub$$
The possible maps are shown in figure 3.  We have used the translational
invariance on the \ws\ to fix one \ram\ point to the origin.  The first
six figures lead to a factor of $6Ag^2$, coming from the integration
of the second \ram\ point over the surface.  (The integration in these figures
is over one cover of the target space).  There are two maps for
each surface shown in figure 1, coming from the two possible ways
to draw the branch cuts.  (There are two and not four since we have
distinguished the points by fixing one to the origin.)

\vskip.2in
{\epsfxsize=6.5in\epsfbox{qcdfig3.eps}}

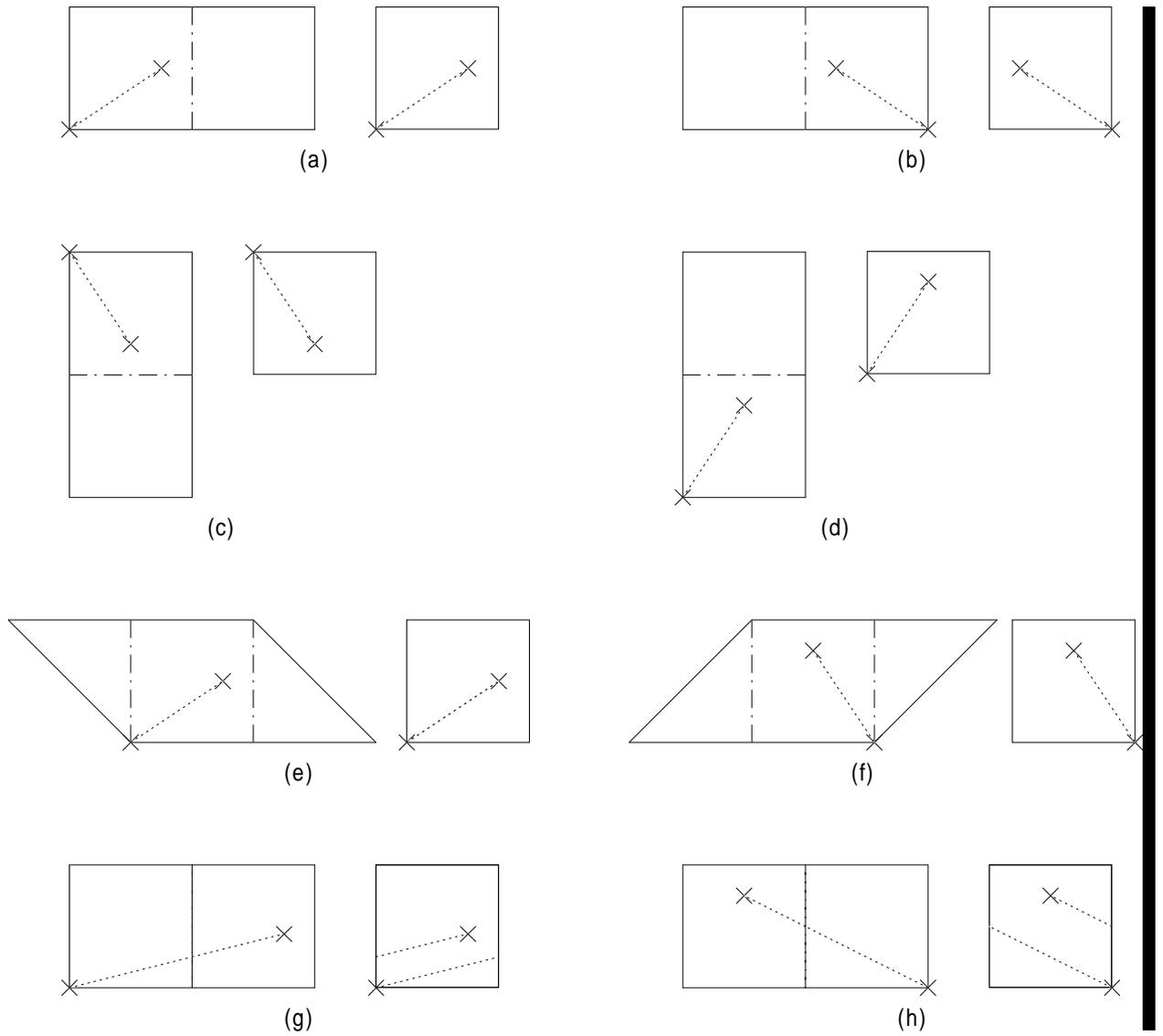
\figcap{3}{Triple covered surfaces with two \ram\ points.  The short dashed
lines are the cuts connceting the two surfaces.  The parallelograms have
periodic boundary conditions.}
\vfill
\eject

\noindent
The last two surfaces in figure 3 contain branch cuts
that wrap more than once around the cycles of the smaller sheet.
For these surfaces
it is only necessary to consider the double covered surface in figure 1a.
It turns out that maps of this type that use the other two surfaces
 in figure 1 are equivalent to the first.
To see this, one can break up the world-sheet into separate
regions and show that the different regions are connected to each other
in the same way for either of the mappings.
This is illustrated in figure 4, where we compare maps containing
the double covered surfaces
pictured in figure 1.
By examining the eight regions in figure 4a
with those in 4b and 4c, one finds that the three surfaces are identical.
(The reader is encouraged to verify this by tracing closed loops around the
surfaces.)
Therefore, there is a factor of $2Ag^2$ from the last two maps in figure 3,
and hence the total sum of factors agrees with \threedoub.

\vskip.2in
{\epsfxsize=6.5in\epsfbox{qcdfig4.eps}}

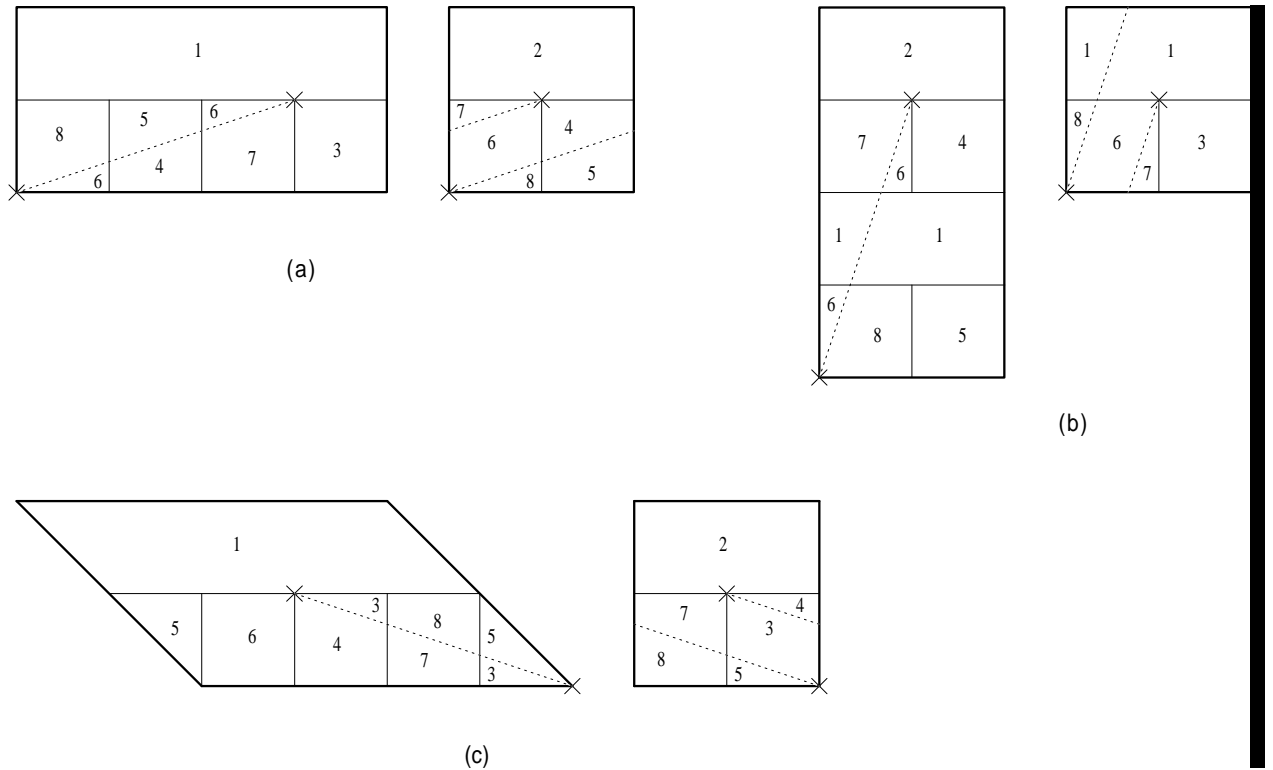
\figcap{4}{Identical triple covered surfaces with two \ram\ points.}


Finally, consider the last term in $\CR$, $-N(n^2/N^2)/2$.  This term appears
to be associated with small handles on the surface and with pinched tubes
connecting different parts of the \ws.  It is convenient to rewrite
$n^2/2N^2$ as
$${n\over 2N^2}+{n(n-1)\over2N^2}.\eqn\rwnsq$$
It is then consistent to say that the first term in \rwnsq\ is related to small
handles and the second term is related to pinched tubes.
For every small handle on the surface, the genus
is increased by one, thus one expects every handle to come with a factor of
$1/N^2$.  Furthermore, the position of the handle needs to be integrated over,
since each position corresponds to a different surface.
Therefore, each handle has
a factor of $nAg^2$, the area of the \ws\ using the \ts\ space metric.
If the handles are infinitesmally small, then their positions are the {\it
only}
moduli. If the handle has finite length, it would have two points
associated with it, corresponding to the points where the handle
is attached to the surface.  In this case, one would integrate over
both points, but with a factor of $1/2$, since the ends
of the handle are indistinguishable.  By shrinking the length of the handle,
the
two points coalesce into one, but the factor of $1/2$ remains.
Finally, interchanging the positions of two handles gives back the same \ws.
Hence, for $n_h$ handles there is a symmetry factor of $1/n_h!$.  Therefore, if
these small handles exist, then every term in the free energy that comes from
a surface that covers the \ts\ $n$ times, should be multiplied by
$\exp(Ag^2n/2N^2)$.  This is precisely the contribution from the first term
in \rwnsq.

If one allows infinitesimally small handles, then consistency requires
that there be infinitesimally small
tubes connecting different points of the \ws.  These points on the \ws\ must
map to the same point on the \ts.  Each tube should come with a factor of
$1/N^2$, since the genus will be increased by one.  Therefore, if we consider
a not necessarily connected surface with area $nA$, and then put in a tube,
this will lead to a factor of $Ag^2n(n-1)/2N^2$, where $Ag^2$ is from the
integration over the \ts, and $n(n-1)/2$ comes from choosing two of the
$n$ sections of the \ws. (By section, we mean a part of the surface that
covers the \ts\ exactly once.)
Again, interchanging two tubes leaves the surface
invariant.  Therefore, if these pinched tubes exist, then
surfaces of area $nA$ also come with a factor of $\exp(Ag^2n(n-1)/2N^2)$
in the partition function.
This agrees with the second part of \rwnsq.

Figure 5 shows such a tube connecting two parts of a \ws.  By examining the
figure, one notes that actually the surface has folded back onto itself.
But the fold takes place at a single point, not along an extensive line.
Hence, the constraints on the \ws\ should read that finite length folds
are suppressed.

We close this section by noting that for the gauge group $U(N)$, the
quadratic casimir is

{\epsfxsize=6.0in\epsfbox{qcdfig5.eps}}
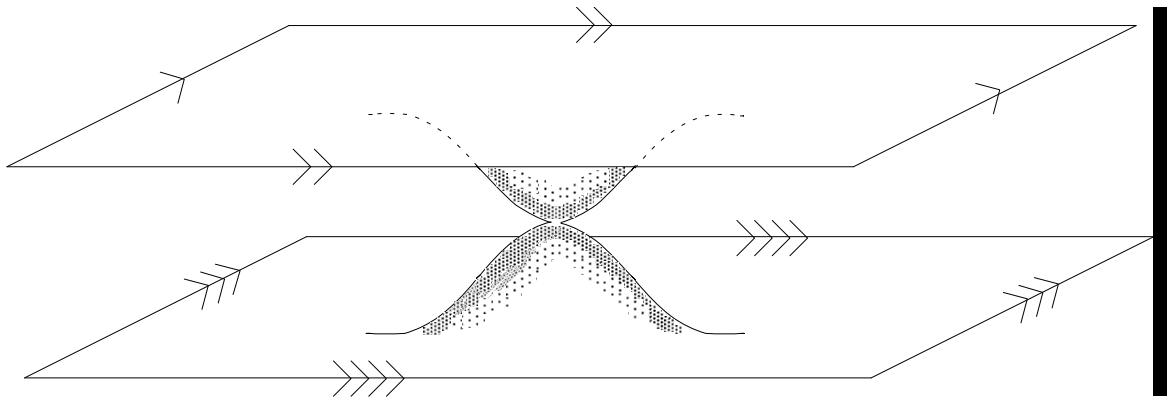
\figcap{5}{Two tori joined by a pinched off tube.  The arrows indicate which
edges are identified on the world sheet.  Note that the surface folds back on
itself at the pinch.}

\noindent
given by
$$\CR={N\over2}(n+{\tn\over N}).\eqn\Uquadcas$$
Hence the string theory corresponding to this version of QCD would have
\ram\ points but not small handles.


\chapter{Discussion}

To summarize, we have given a string interpretation
for all terms found in the perturbative
expansion of the QCD partition function given in \partfun.
The free energy is given by
a sum over maps from the \ws\ modded out by \diffs.
Unlike the Polyakov string, there is no integration over \ws\ metrics.
The free energy contains a sum over branched surfaces with small handles,
which are inequivalent under \diffs.
We have verified the consistency of this interpretation for the lower
order terms in the expansion.

We close with a few remarks.
We first note that there is actually a natural way to suppress the folds.
This is accomplished
by introducing an extrinsic curvature term in the action.  For a
two-dimensional
surface mapped into another two dimensional surface, this is given by
$$\int d^2\xi \eta^{ab}\partial_an\partial_bn.\eqn\excurv$$
$n$ is basically the normal ``vector'' to the surface,
which in this case is $n=\pm1$.  At a fold, the derivative normal to the fold
on the surface is a delta function.  Hence the integral in \excurv\ is
$L\delta(0)$, where $L$ is the length of all folds on the surface.  Hence
finite length folds will be suppressed by this term.  This suggests that
an analogous term might appear in the four dimensional case as well,
although in this case, the bending of the surface will lead to finite results.

Our second remark concerns nonperturbative effects.
The partition function in \partfun\ is an even function of $1/N$ only up
to terms with $N$ boxes in the tableaux.  The fact that the entire sum
will not be an even function is then a nonperturbative result.  This then
complies with Shenker's observation that nonperturbative string effects
are on the order of $\exp(-1/g_s)$ and not $\exp(-1/g_s^2)$.  If the latter
case were true then the complete partition function would be an even function
of $1/N$.  Another way to understand what determines the order of the
nonperturbative effects is to realize that there are corrections to
the perturbative sum when the number of coverings of the \ts\ is
$N$ or greater.  This is because tableaux have been summed over that
don't correspond to physical representations of $SU(N)$.
Hence, the free energy will have correction terms of the form
$\exp(-Ag^2N/2)$ multiplied by moduli factors.

Finally, while it is difficult to verify that the free energy gives
a single counting of branched maps into the torus,  it is not too
hard to actually calculate the terms in the free energy.
Hence one can turn this
around and simply postulate what the distinct maps are by reading off the terms
in the free energy.  Using the symbolic manipulator program Maple, we were
able to calculate such terms for surfaces that cover the \ts\ up to 10
times and with as many as six \ram\ points.
These results are shown in table 1.
This is then another instance where quantum field theory can be
used to explore questions in geometry.
\vskip.2in

\vbox{\tabskip=0pt \offinterlineskip
\def\tablerule{\noalign{\hrule}
\omit&height2pt&\omit&&\omit&&\omit&&\omit&\cr}
\halign to350pt{\strut#&\vrule#\tabskip=1em plus2em&
\hfil#& \vrule#& \hfil#& \vrule#& \hfil#& \vrule#&
\hfil#& \vrule#\tabskip=0pt\cr\tablerule
&&Covers&&
2 Ram. Pts.&&
4 Ram. Pts.&&
6 Ram. Pts.&\cr\tablerule
&& 1&&0&&0&&0&\cr\tablerule
&& 2&&$2q^2$&&$(1/12)q^4$&&$(1/360)q^6$&\cr\tablerule
&& 3&&$8q^2$&&$(20/3)q^4$&&$(91/45)q^6$&\cr\tablerule
&& 4&&$30q^2$&&$102q^4$&&$(383/3)q^6$&\cr\tablerule
&& 5&&$80q^2$&&$(2288/3)q^4$&&$(24140/9)q^6$&\cr\tablerule
&& 6&&$180q^2$&&$3773q^4$&&$(180331/6)q^6$&\cr\tablerule
&& 7&&$336q^2$&&$14232q^4$&&$(3349714/15)q^6$&\cr\tablerule
&& 8&&$620q^2$&&$(133616/3)q^4$&&$(11174816/9)q^6$&\cr\tablerule
&& 9&&$960q^2$&&$119904q^4$&&$5558312q^6$&\cr\tablerule
&&10&&$1590q^2$&&$(584517/2)q^4$&&$(252779965/12)q^6$&\cr\noalign{\hrule}
}}
\tablecap{1}{Multiplicative factors from integrations over positions of
the \ram\ points.  $q=Ag^2$}

\noindent {\it Note added}:  After this paper was completed, we learned
that Gross and Taylor were able to prove that the free energy counts
the number of independent maps (without branch points or small handles)
for any number of coverings.  They also
extended this to target spaces with genus $G>1$. [\GrTay]

\ack{This research was supported in part by D.O.E. grant DE-AS05-85ER-40518.}
\refout
\end